\documentclass[acmlarge,nonacm]{acmart}
\renewcommand\footnotetextcopyrightpermission[1]{} 

\usepackage{pifont} 

\usepackage{hyperref}
\usepackage{float}

\def\BibTeX{{\rm B\kern-.05em{\sc i\kern-.025em b}\kern-.08emT\kern-.1667em\lower.7ex\hbox{E}\kern-.125emX}}
    
\begin{document}

%
\title{The centaur programmer - How Kasparov's Advanced Chess spans over to the software development of the future}

%


\author{Pedro Alves}
\affiliation{%
  \institution{Lusófona University}
  \streetaddress{Campo Grande, 376}
  \city{Lisbon}
  \country{Portugal}
  \postcode{1700-097}
}
\email{pedro.alves@ulusofona.pt}

\author{Bruno Pereira Cipriano}
\affiliation{%
  \institution{Lusófona University}
  \streetaddress{Campo Grande, 376}
  \city{Lisbon}
  \country{Portugal}
  \postcode{1700-097}
}
\email{bcipriano@ulusofona.pt}

%

%
\begin{abstract}
We introduce the idea of Centaur Programmer, based on the premise that a collaborative approach between humans and AI will be more effective than AI alone, as demonstrated in centaur chess tournaments where mixed teams of humans and AI beat sole computers.

The paper introduces several collaboration models for programming alongside an AI, including the guidance model, the sketch model, and the inverted control model, and suggests that universities should prepare future programmers for a more efficient and productive programming environment augmented with AI.

We hope to contribute to the important discussion about the diverse ways whereby humans and AI can work together in programming in the next decade, how universities should handle these changes and some legal implications surrounding this topic.
\end{abstract}

%
%


%
\keywords{Position Paper, Artificial Intelligence, Programming}

%

%
\maketitle

\section{Introduction}

The victory of a computer (Deep Blue) against Gary Kasparov, the world chess champion, in 1997 was a milestone in AI history and led many to predict the substitution of humans by machines in a wide range of activities \cite{hassabis_artificial_2017}. The digital revolution had been dismissing repetitive, highly standardized tasks with little need for intellectual skills, but now it seemed that even highly complex tasks could be performed by machines, at low cost and high efficiency.

One such task is programming itself. While many workers were already concerned about maintaining their jobs, this was not on the minds of the millions of programmers fueling the digital revolution. After all, someone has to create the systems that support this revolution. Recently, LLM-based tools (e.g., GPT-3) have shown great effectiveness in programming tasks, being able to solve exercises at the level of an average student in a Computer Engineering course \cite{finnie-ansley_robots_2022}. New versions promise to improve this performance \cite{openai_gpt-4_2023}. Is this the Deep Blue vs. Gary Kasparov moment of programming? Will even programmers themselves be replaced by AI?

We cannot predict the future, but we can learn from the past. Returning to the world of chess, after Kasparov's defeat, there were several experiments with competitions involving humans and AI, which were called "Centaur Chess"\footnote{Also known as "Advanced Chess"} in analogy to the mythological figure of the Centaur that was half human, half horse. From these experiments, one, in particular, brought interesting findings. In 2005, there was a chess tournament in which all kinds of entities could compete: computers, chess champions, and mixed teams composed of humans and computers. It was already expected that computers would beat chess champions, but what was not expected was that mixed teams would beat computers. In other words, the combination of human + AI seemed to be more effective than AI alone. The idea of the Centaur as a model for future work was starting to take shape \cite{case_how_2018}.

In this paper, we argue that something similar will happen in programming. A team formed by programmers (humans) and AI will be more effective than AI alone, hence the term Centaur Programmer.

We propose several models under which this collaboration may occur, show how universities may adapt to this new reality, draw attention to legal and ethical implications and finish with some conclusions about what the future may hold for this topic.

\section{Collaboration Models}

The collaboration between human and AI has already been successfully proven in other areas. Maurice Conti demonstrated that, with the help of AI, he was able to design very efficient equipment that would never have seen the light of day if the teams were purely made up of humans or AI \cite{heaven_designer_2018}. The process was simple: humans defined objectives and constraints related to a certain problem, and AI evolved possible solutions with constant validation from humans.

This collaboration model, which we can call the "guidance model," is one of the most interesting possibilities also for the programming area. The programmer starts the process, indicating to the AI what the objectives and assumptions are, and the AI responds with solutions. With the help of the programmer, these solutions evolve until they reach the desired result. The final solution may or may not be better than a solution purely imagined by the programmer, but it will probably be achieved much more quickly. For now, we are mainly talking about efficiency gains rather than new solutions. But AI technologies are evolving rapidly, and, in the near future, we may actually have better-designed programs with fewer bugs and more efficient if created in a collaborative model.

Other collaboration models may be explored in software development. In the "sketch model," the programmer outlines the program's structure and the AI fills in the gaps. While tools like GitHub Copilot (based on the GPT technology) currently seem to fit this model \cite{peng_impact_2023}, we believe they have a limited role, acting on a function-by-function basis rather than holistically (i.e., "I will create this class because it will be useful in N other parts of the program").

In the "inverted control model," the communication direction is reversed: the AI asks the programmer what they intend to do until it objectively understands the goal and constraints before implementation.

All of these models require a rethinking of the software development cycle. Programmers are already accustomed to "boosting" their efficiency through a "query model" where they simply ask questions to the computer (e.g., Stack Overflow). The initial reaction to tools like ChatGPT has been to transpose this model - when information is lacking, the programmer asks ChatGPT in a similar manner to the query they would make on Google or Stack Overflow. But is this the most efficient model? In this model, the programmer still does most of the typing, wasting the potential gains that a true programmer-centaur can have.

\section{The role of universities}

If these new models are to profoundly transform the software development industry, that transformation must begin in universities, where future programmers are being educated. Universities should start exploring and integrating these new collaboration models into their software development curricula to prepare the future workforce for a more efficient and productive programming environment. It is not enough to prepare future programmers, they must prepare future centaur programmers.

We believe that this transformation will be gradual and experimental. In the first phase, we will have the model of the "virtual tutor", where the student uses AI to help them learn. This tutor can explain to the student what a certain piece of code does or what a particular error means. It can help them verify if a certain solution is correct or propose improvements to the code developed by the student. It can even convert a program to different programming languages. In any case, we are talking about a learning-centered process - the student wants to know more and AI helps them in this process, complementing the role of the teacher. After all, the idea of having a personal tutor dedicated to each student would have seemed utopian until recently due to the costs it would entail. Not anymore with AI.

But the "virtual tutor" model, while important, has nothing to do with the idea of the centaur-programmer, as the objectives are different. The centaur-programmers aim to augment their ability through collaboration with AI. This can translate into greater efficiency (producing solutions more quickly than they would without AI) and/or greater quality (producing solutions of better quality than they would alone). Returning to the initial concept of "centaur chess", the goal was to win the game, not so much to learn to play chess.

Therefore, a second phase will emerge in which students actually learn to augment their abilities with the help of AI. The analogy of the scientific calculator has been widely used and seems appropriate. The calculator alone is of little use because it does not know which calculations to perform. And the human alone is inefficient because it takes too long to perform certain calculations. A human equipped with a calculator is, in fact, a centaur that uses a collaboration model accepted in the academic world.

Is this analogy valid? A calculator simply reproduces the same mechanical process as a student performing mathematical operations (albeit much more quickly). However, tools like ChatGPT can create new content from a set of interactions. Students can abuse these tools to solve complete programming exercises without actually knowing how to program. This has even led several universities to simply ban/prohibit the use of these tools \cite{jimenez_this_2023}. But if the industry is massively adopting the use of these tools due to the potential increase in productivity that results from it, shouldn't universities follow the same path?

Returning to the analogy, there is a key point that remains valid. Both the calculator and ChatGPT (or other forms of AI) are dependent on a human to produce something useful. In fact, the level of usefulness that can be extracted from these tools depends directly on the ability of the human using them. An advanced mathematics student can certainly use a scientific calculator more effectively and \textit{usefully} than a primary school student - here we use the term "useful" to refer to the ability to solve a real problem rather than just solve academic exercises. The same goes for experienced programmers who will use their analytical and creative ability to create programs that solve real problems. The pieces that make up these programs may be developed by AI, but knowing which pieces to build and how to connect them will continue to be the exclusive responsibility of the human. If it was already clear that universities should invest more in developing these capabilities, it now becomes even more evident.

\section{Legal and ethical implications}

There has been much discussion about the ethical and legal implications of using AI as a tool, particularly within the subset of AI based on LLMs, as it is trained on large quantities of information generated by humans. On the one hand, the unauthorized use of content raises legal questions. On the other hand, it introduces biases that replicate patterns of behavior in society and may lead to racism, xenophobia, or other types of discrimination.

Here, too, the figure of the centaur programmer is important as it assumes prior (human) validation of the content produced by AI. This validation derives from an assumption that seems essential to us: the content produced by a collaboration between humans and AI should ultimately be the sole responsibility of humans. For example, humans may prefer to use LLMs trained on a dataset previously curated by them to ensure that: (1) there are no legal problems with the unauthorized use of content; (2) there is no incorrect or biased information feeding the training engine.

In this context, Universities also play a critical role in developing and openly discussing the ethical guidelines for the use of AI technologies. It makes sense for universities to host and train their own LLMs, based on data produced by the university itself or previously curated by it, making sure it adheres to the previously defined guidelines.
\section{Conclusion}

Centaur-programmers are coming. They will be humans able to augment their programming abilities by collaborating with AI. They will accomplish the software requirements much faster and with better quality than their sole human counterparts. They will be trained by their university to explore different collaboration models with AI, such as the guidance and the sketch model. They will also have a strong critical spirit and highly ethical values that will act as a safeguard to the content produced by AI. Because, in the end, the code may have been created by AI but its the human programmer who will take responsibility for it.

\bibliographystyle{ACM-Reference-Format}
\bibliography{bibliography}

%

\end{document}